\documentclass[conference]{IEEEtran}

\ifCLASSINFOpdf
\else
\fi

\usepackage[usenames,dvipsnames]{xcolor}
\usepackage{amsmath,amssymb,amsfonts,tikz,pgfplots,bm}
\usepackage{cite,booktabs,tabularx}
\usepackage{algorithmic}
\usepackage{graphicx}
\usepackage{textcomp}
\usepackage{verbatim}
\usepackage{amsmath}
\usepackage[ruled,vlined]{algorithm2e}
\usepackage{nicefrac}
\usepackage[inline]{enumitem}
\usepackage{array}
\usepackage{balance}

\newcommand{\len}{\ensuremath{N}}
\newcommand{\leni}{\ensuremath{n}}
\newcommand{\mem}{\ensuremath{m}}
\newcommand{\dimi}{\ensuremath{k}}
\newcommand{\dimibit}{\ensuremath{k}}
\newcommand{\leno}{\ensuremath{N_\mathsf{o}}}
\newcommand{\dimo}{\ensuremath{K}}
\newcommand{\numS}{\ensuremath{M}}
\newcommand{\bl}{\leno+m}

\newcommand{\rateo}{\ensuremath{R_\mathsf{o}}}
\newcommand{\ratei}{\ensuremath{R_\mathsf{i}}}

\newcommand{\id}{\ensuremath{i}}

\newcommand{\D}{\mathsf{D}}
\newcommand{\I}{\mathsf{I}}
\newcommand{\T}{\mathsf{T}}
\renewcommand{\S}{\mathsf{S}}

\let\vv\v

\renewcommand{\u}{\ensuremath{\boldsymbol{u}}}
\newcommand{\w}{\ensuremath{\boldsymbol{w}}}

\newcommand{\x}{\ensuremath{\boldsymbol{x}}}
\newcommand{\y}{\ensuremath{\boldsymbol{y}}}
\renewcommand{\v}{\ensuremath{\boldsymbol{v}}}

\newcommand{\V}{\ensuremath{V}}

\newcommand{\Al}{\ensuremath{\Sigma}}

\usetikzlibrary{positioning,calc}
\tikzstyle{block} = [rectangle, draw, text centered, rounded corners, minimum height=1.5em,fill=blue!7!]

\tikzstyle{round} = [circle, draw, text centered, minimum height=1.5em, fill=blue!7!]

\renewcommand{\baselinestretch}{0.965} 

\newcommand{\field}[1]{\ensuremath{\mathbb{F}_{#1}}}

\newcommand{\outq}{\ensuremath{q_\mathsf{o}}}

\newcommand{\olen}{\ensuremath{N_\mathsf{o}}}

\newcommand{\ocw}{\ensuremath{\boldsymbol{w}}}

\newcommand{\tv}{\ensuremath{t}}

\newcommand{\concat}{\ensuremath{\mathbin\Vert}}

\newcommand{\Ebb}{\ensuremath{\mathbb{E}}}

\definecolor{GreenB}     {rgb}{0.37,0.57,0.42}

\definecolor{ColorIBDDSR}{rgb}{1,0,1}
\definecolor{mycolor4}{rgb}{0.07843,0.16863,0.54902}
\definecolor{ColoriBDD(ideal)}{rgb}{0.85098, 0.32941, 0.10196}
\definecolor{bazaar}{rgb}{0.6, 0.47, 0.48}
\definecolor{ColorIGMDDSR}{rgb}{0,0.49804,0}

\newcommand{\Pfe}{P_{\mathsf f}(e)}
\newcommand{\bsw}{\boldsymbol{\mathsf{w}}}
\newcommand{\bsy}{\boldsymbol{\mathsf{y}}}

\begin{document}

\title{Finite Blocklength Performance Bound\\for the DNA Storage Channel}

\author{\IEEEauthorblockN{{\bf Issam Maarouf}\IEEEauthorrefmark{1}, {\bf Gianluigi Liva}\IEEEauthorrefmark{2}, {\bf Eirik Rosnes}\IEEEauthorrefmark{1}, and {\bf Alexandre Graell i Amat}\IEEEauthorrefmark{3}}

\IEEEauthorblockA{
	\IEEEauthorrefmark{1}Simula UiB, N-5006  Bergen,  Norway
}

\IEEEauthorblockA{
	\IEEEauthorrefmark{2}Institute of Communications and Navigation, German Aerospace Center, 82234 Weßling, Germany
}

\IEEEauthorblockA{
	\IEEEauthorrefmark{3}Department of Electrical Engineering, Chalmers University of Technology, SE-41296 Gothenburg, Sweden
}%
}

\maketitle

\begin{abstract}
We present a finite blocklength performance bound for a DNA storage channel with insertions, deletions, and substitutions. The considered bound\textemdash the dependency testing (DT) bound, introduced by Polyanskiy \emph{et al.} in 2010\textemdash, provides an upper bound on the achievable frame error probability and can be used to benchmark coding schemes in the practical short-to-medium blocklength regime. In particular, we consider a concatenated coding scheme where an inner synchronization code deals with insertions and deletions and the outer code corrects remaining (mostly substitution) errors. The bound depends on the inner synchronization code. Thus, it allows to guide its choice. We then consider low-density parity-check codes for the outer code, which we optimize based on extrinsic information transfer charts. Our optimized coding schemes achieve a normalized rate of $87\%$ to $97\%$ with respect to the DT bound for code lengths up to $2000$ DNA symbols for a frame error probability of $10^{-3}$ and code rate $\nicefrac{1}{2}$.
\end{abstract}

\IEEEpeerreviewmaketitle

\section{Introduction} \label{sec:introduction}
Using deoxyribonucleic acid (DNA) as a medium to store data is seen as the next frontier of data storage, providing unprecedented durability and density. Several experiments  have already  demonstrated the viability of DNA-based data storage, see, e.g., \cite{yazdi_portable_2017,organick_random_2018}. %

The DNA storage channel is impaired by  insertions, deletions, and substitutions (IDSs) arising from the synthesis and sequencing of DNA sequences\cite{church_next-generation_2012}. %
Hence, reliable storage of data in DNA requires the use of error-correcting codes. Designing a code that handles IDS errors jointly is, however, a daunting task. Davey and MacKay \cite{davey_reliable_2001} proposed a clever solution to this problem by introducing a serially-concatenated coding scheme (for the binary IDS channel) in which the inner code, called \emph{synchronization} code, deals with insertions and deletions, and the outer code (a low-density parity-check (LDPC) code  in \cite{davey_reliable_2001}) corrects remaining errors, mostly in the form of substitutions.

The literature on coding for DNA storage is abundant. Most works consider a very small number of deletions and/or insertions\textemdash i.e., an adversarial channel\textemdash and a single DNA strand. In DNA-based storage, however, errors occur probabilistically and can be substantial, and the synthesis and sequencing processes  result in multiple (noisy) copies of the same DNA strand.
The authors in \cite{maarouf2021concatenated} were the first to introduce decoding algorithms for coding schemes exploiting multiple reads of the DNA sequence. The work \cite{maarouf2021concatenated} was followed by \cite{Srinivasavaradhan2021TrellisBMA}. 

The works \cite{maarouf2021concatenated} and \cite{Srinivasavaradhan2021TrellisBMA} also provided achievable information rates, which give insight into the performance of  coding schemes with very large blocklengths. However, current DNA storage technology only supports the synthesis and sequencing of short-to-medium-length DNA strands, in the range of $100$-$2000$ DNA symbols. Therefore,  performance bounds for the finite blocklength regime would be  more informative for the DNA channel.
To the best of our knowledge, no finite blocklength performance bounds for the DNA storage channel (and IDS channels in general) exist in the literature.

In this paper,  we   provide a finite blocklength performance bound for a DNA  storage channel with IDS errors. Particularly, we consider the dependency testing (DT) bound  \cite{DT+RCUsbounds} based on the \emph{random coding} principle, which gives an upper bound on the frame error probability achievable over the DNA storage channel. The bound is tailored to a concatenated coding scheme that uses an inner synchronization code and  depends on the inner code. Hence, it can be used as a handy tool to optimize the inner synchronization code for the finite blocklength regime. Further, the bound provides a benchmark to compare coding schemes for DNA storage in the practical short-to-medium blocklength regime.
We also consider the optimization of an outer  LDPC code for a given inner code using extrinsic information transfer (EXIT) charts, and show that an optimized concatenated coding scheme  achieves a normalized rate of $87\%$ to $97\%$ with respect to the DT bound for a frame error probability of $10^{-3}$ and code rate $\nicefrac{1}{2}$, depending on the sequence length. These values are similar to those of state-of-the-art coding schemes for simpler memoryless channels (such as the Gaussian channel and the binary symmetric channel), highlighting that the scheme in \cite{maarouf2021concatenated} achieves excellent performance for the DNA storage channel in the short-to-medium blocklength regime.

\section{System Model} \label{sec:system}
\subsection{Channel Model}

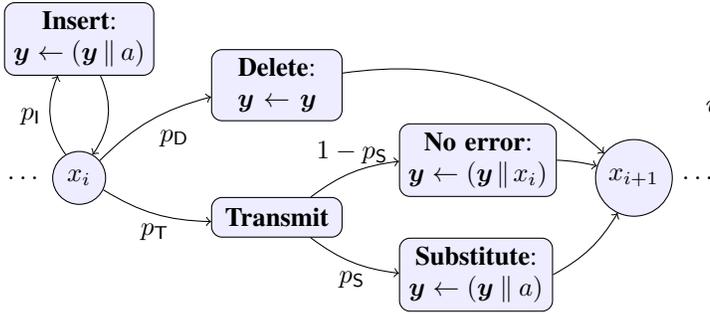
\begin{figure}[t]
	\centering
		\begin{tikzpicture}
		\node[round] (xi) {$x_i$};
		\node[left=0cm of xi] {$\dots$};
		
		\node[block, above=1cm of xi, text width=1.75cm] (ins) {{\bf Insert}: $\y \leftarrow (\y \concat a)$};
		
		\node[block, above right=.5cm and 1.5cm of xi, text width=1.5cm] (del) {{\bf Delete}: $\y \leftarrow \y$};
		
		\node[block, below right=.0cm and 1.5cm of xi, text width=1.5cm] (tr) {{\bf Transmit}};
		
		\node[block, above right=.0cm and .75cm of tr, text width=1.85cm] (noerr) {{\bf No error}: $\y \leftarrow (\y \concat x_i)$};
		\node[block, below right=.0cm and .75cm of tr, text width=1.8cm] (sub) {{\bf Substitute}: $\y \leftarrow (\y \concat a)$};
		
		\node[round, right=6.5cm of xi] (xip1) {$x_{i+1}$};
		\node[right=0cm of xip1] {$\dots$};
		
		\draw[->] (xi) to [bend left] node[left] {$p_\I$} (ins);
		\draw[->] (ins) to [bend left] (xi);
		
		\draw[->] (xi) to [bend left=15] node[below right] {$p_\D$} (del);
		
		\draw[->] (xi) to [bend right=15] node[below] {$p_\T$} (tr);
		
		\draw[->] (tr) to [bend right=15] node[below] {$p_\S$} (sub);
		\draw[->] (tr) to [bend left=15] node[above] {$1-p_\S$} (noerr);
		
		\draw[->] (del) to [bend left=25] (xip1);
		
		\draw[->] (noerr.east) to [bend left=5] (xip1);
		\draw[->] (sub.east) to [bend right=15] (xip1);
	\end{tikzpicture}
	\vspace{-3ex}
	\caption{State-based representation of the DNA storage channel with IDS errors.}
	\label{fig:ids:channel}
	\vspace{-2ex}
\end{figure}

We consider the widely-used simplified channel model depicted in Fig.~\ref{fig:ids:channel} \cite{davey_reliable_2001, briffa_improved_2010} for the DNA storage channel, where IDS errors are independent and identically distributed.
Let $\x = (x_1,\ldots, x_\len)$, $x_i \in \Al _q = \{0,1,\dots,q-1\}$,\footnote{For the DNA storage channel, $q=4$. However, we use $q$ for the sake of generality.} be the information DNA sequence of length $N$ to be transmitted over the channel. The sequence can be viewed as a queue of symbols, where each symbol $x_i$ is successively transmitted over the channel. The received sequence  $\y = (y_1,\ldots, y_{\len'})$, where 
$N'$ may  be different to $N$ due to insertions and deletions, is generated state by state and is obtained as follows. Assume $x_i$ is next in queue to be transmitted over the channel. The channel enters state $x_i$ where three events may occur: i) an insertion event, with probability $p_\I$, where a random symbol $a \in \Al_q$ is appended to  $\y$ instead of $x_i$. In this case, $x_i$ remains in the queue and the channel returns to state $x_i$; ii) a deletion event, with probability $p_\D$, where symbol $x_i$ is deleted from the queue. In this case, nothing is appended to $\y$, the next symbol $x_{i+1}$ is enqueued, and the channel enters state $x_{i+1}$; iii) a transmission event, with probability $p_\T = 1-p_\I -p_\D$, where $x_i$ is transmitted. In this case, the symbol $x_i$ is either received with no error with probability $1 - p_\S$ or in error with probability $p_\S$, in which case $x_i$ is substituted  by a random symbol $a \neq x_i$. In either case, the next symbol $x_{i+1}$ is enqueued, and the channel enters state $x_{i+1}$. The process finishes when the last symbol $x_{\len}$ leaves the queue. The channel output is $\y$. 
 
 The difference $\len - \len'$ is referred to as the \emph{drift} \cite{davey_reliable_2001} at the end of the transmitted sequence. We can also define a drift  for each symbol $x_i$ to be transmitted, or each time instant $i$. Formally, the \emph{symbol-level}  drift $d^\text{sym}_i$, $0 \leq i < \len$, is defined as the difference between the number of insertions and the number of deletions that occurred before symbol $x_{i+1}$ is enqueued, while $d^\text{sym}_\len$ is defined as the number of insertions minus deletions that occurred after the last symbol
$x_{N}$ has been transmitted.

Finally, we model the multiple reads of a DNA sequence resulting from the synthesis and sequencing processes as transmitting the DNA sequence $\x$ over $M$  parallel and independent IDS channels, see Fig.~\ref{fig:system:model},
resulting in the received sequences $\y_1,\dots,\y_{\numS}$. %

\subsection{Coding Scheme}

\begin{figure}[t]
	\centering
		\begin{tikzpicture}

		\node (u) {$\u$};
		
		\node[block, right=0.35cm of u,text width=1.5cm] (out) {\bf Outer code};
		
		\node[block, below=1.2cm of out,text width=1.8cm] (inn) {\bf Inner code w. offset};
		
		\node[circle, fill, right=0.55cm of inn,inner sep=.035cm] (circ) {};
		
		\draw (inn) -- node[above,pos=.45] {$\x$} (circ);

	    \node[block, above right = 1.45cm and 1.00cm of inn, text width=1.3cm] (c1) {\bf IDS channel};
		\node[block, above right = -0.05cm and 1.00cm of inn, text width=1.3cm] (c2) {\bf IDS channel};
		\node[block, below right = 0.05cm and 1.00cm of inn, text width=1.3cm] (cA) {\bf IDS channel};
		\node[above = .15cm of cA] (cd) {$\vdots$};
		
		\node[block,right=3.5cm of inn, text width=1.8cm] (inndec) {\bf Inner decoder};
		
		\node[block,above=1.15cm of inndec, text width=1.5cm] (outdec) {\bf Outer decoder};
		
		\node [right=0.3cm of outdec](uh) {$\hat{\u}$};
		
		\draw[->] (u) -- (out);
		\draw[->] (out) -- node[left] {$\w$} (inn);

		\draw[->] (circ) |- (c1.west);
		\draw[->] (circ) -- node[above]{}+(0.1,0) |- ($(c2.west) + (0,0)$);
		\draw[->] (circ) |- (cA.west);

		\draw [dashed] ($(c1)+(-1.35,1.1)$) rectangle ($(cA)+(1.45,-0.8)$);
		\node[anchor=west] at ($(c1)+(-0.7,.9)$) {Channel};

		\draw[->] (inndec) -- node[right] {$p(w_i|\y)$} (outdec);
		\draw[->, dashed] ($(outdec.south) + (-0.2,0)$) -- ($(inndec.north) + (-0.2,0)$);
		\draw[->] (outdec) -- (uh);

        \draw[->] (c1.east) -- node[above] {$\y_1$}+(0.5,0) |- ($(inndec.west) + (0,0.25)$);
        \draw[->] (c2.east) -- node[above] {$\y_2$}+(0.4,0) |- ($(inndec.west) + (0,0)$);
	    \draw[->] (cA.east) -- node[above] {$\y_M$}+(0.5,0) |- ($(inndec.west) + (0,-0.25)$);

	\end{tikzpicture}
	\vspace{-3ex}
	\caption{Block diagram of the encoder and decoder for the DNA storage channel. The DNA storage channel is modeled as multiple reads of the DNA strand transmitted over  parallel IDS channels: the channel depicted in Fig.~\ref{fig:ids:channel} is fed $\numS$ times with the DNA sequence $\x$. Here, $\y = (\y_1,\dots, \y_{\numS})$.}
	\label{fig:system:model}
	\vspace{-2ex}
\end{figure}
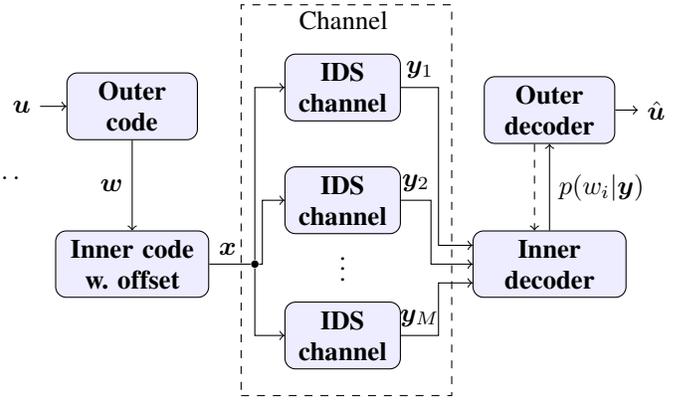

We consider a concatenated coding scheme with an inner synchronization code depicted  in Fig.~\ref{fig:system:model}. First, the information sequence $\u = (u_1, \ldots , u_\dimo)$,   $u_i \in \field{\outq}$, is encoded by an $[\leno,\dimo]_{\outq}$ outer code to produce a codeword $\ocw = (w_1,\dots,w_{\leno})$, $w_i \in \field{{\outq}}$, where  $\field{{\outq}}$ is a binary field extension with $\outq = 2^\dimi$. The codeword $\w$ is then encoded by an inner synchronization code. Here, we consider block  and convolutional codes for the inner code. We denote the block code by $[\leni,\dimi, \tv]_q$, where $\leni$ and $\dimi$ are the length and dimension of the code, respectively, and $\tv$ represents the number of different codebooks that are used (see~\cite{maarouf2021concatenated} for details). Furthermore, the convolutional code is denoted by $(\leni,\dimi, \mem)_q$, where $\mem$ is the number of memory elements. For simplicity, in the rest of the paper we will consider an inner convolutional code for notations and equations.  We denote the  codeword of the inner code by $\v = (v_1,\ldots,v_\len)$, $v_i \in \Sigma_q$, which is of length  $\len = (\olen + \mem) \leni$ due to termination of the convolutional code. Finally, a pseudo-random offset sequence is optionally added to $\v$ before transmission for synchronization purposes \cite{davey_reliable_2001,Buttigieg_CodebookAM_2011}, resulting in the sequence $\x = (x_1,\ldots,x_\len)$. (A detailed explanation of the role of the random sequence in maintaining synchronization and aiding the decoding of  the inner code is given in \cite{maarouf2021concatenated}.) The DNA sequence $\x$ is finally stored in the DNA medium.

The coding scheme rate is measured in bits per DNA symbol (i.e., per nucleotide) and is given by $R = \rateo \ratei = \nicefrac{\dimo \dimibit}{\len}$, where  $\rateo = \nicefrac{\dimo}{\leno}$ and $\ratei = \nicefrac{\olen \dimibit}{\len}$ are the rates of the outer and inner code, respectively. As we will be only concerned with the drift at time instances that are multiples of $n$, we define the  shorthand $d_i \triangleq d^\text{sym}_{i  n}$. Note that  $d_0 = 0$ and $d_{\bl} = N'-N$, both known to the receiver.

To recover the information sequence $\u$, the inner decoder uses the (noisy) multiple reads $\y_1,\dots,\y_{\numS}$ of the DNA sequence $\x$ to compute (approximate)  a posteriori probabilities (APPs) for the symbols in $\w$. These APPs are then fed to the outer decoder, which decides on the decoded sequence $\hat{\u}$. Furthermore, we can also iterate between the inner and outer decoder, exchanging extrinsic information between them, which is referred to as \emph{turbo decoding} in the literature.

\section{Bound on the Finite Blocklength Performance} \label{sec:info_density}

In this section, we provide an upper bound to the frame error probability, denoted by $\Pfe$, achievable over the DNA storage channel in the finite blocklength regime. In particular, we consider the DT bound  \cite{DT+RCUsbounds}. The bound we provide is tailored to concatenated coding schemes with an inner synchronization code and depends on  the  inner code. Hence, it can be used to guide its choice and serves as a benchmark to compare coding schemes.

The DT bound for the combination of the inner code %
and the DNA storage channel is given by 
\begin{align}
\label{eq:BoundPf}
        \Pfe \leq \;&\Ebb\left[2^{-\left(\id(\bsw;\bsy) - \log_2\frac{\outq^{\leno}-1}{2}\right)^{+}}\right]\,,
\end{align}%
where $(x)^+ \triangleq \max(x,0)$, $\Ebb[\cdot]$ denotes expectation, $\bsy=(\bsy_1,\ldots,\bsy_M)$ for the multiple sequences case, and 
 \begin{align*}
 \id(\bsw;\bsy) \triangleq \log_2\frac{p(\bsy|\bsw)}{p(\bsy)}  %
 \end{align*}
 is the so-called \emph{information density} with expected value equal to the mutual information between $\bsw$ and $\bsy$.\footnote{In order to distinguish between random variables and their realizations, $\bsw$ and $\bsy$ denote  the random variables corresponding to $\w$ and $\y$, respectively.} %
The distribution of the information density $\id(\bsw;\bsy)$  is not known in closed form for the DNA storage channel. %
However, the right-hand-side of \eqref{eq:BoundPf} can be accurately estimated  using the Monte-Carlo approach proposed in \cite{arnold_simulation-based_2006,pfister_achievable_2001}, which exploits concentration properties of Markov chains to estimate
the mutual information between an input process $\w$ and an output process $\y$ via trellis-based simulations. We can then approximate  \eqref{eq:BoundPf} by
\begin{align}
\label{eq:BoundPf2}
        \Pfe \lesssim \; &\frac{1}{\V} \sum_{(\w,\y)}2^{-\left(\id(\w;\y) - (\leno \log_2 \outq-1)\right)^{+}}\,,
\end{align}%
 where $V$ is the number of  pairs $(\w,\y)$ considered in the computation.

In the following, we  show how to {efficiently compute $\id(\w;\y)$ for fixed $\w$ and $\y$. We stress that the values of $\id(\w;\y)$ and, hence, their distribution depends on the choice of the inner code.  
The information density can be written as
\begin{align}
	\id(\w; \y) = -\log_2 p(\w) -\log_2 p(\y) + \log_2 p(\w,\y)\,,
	\label{eq:MIrate}
\end{align}
where the probabilities $p(\w)$, $p(\y)$, and $p(\w, \y)$ can be computed using the forward recursion of the symbol-wise maximum a posteriori  decoding algorithm on the  trellis describing the combination of the inner code and the DNA storage channel \cite{maarouf2021concatenated} (hereafter in this paragraph referred to as simply the inner code for the sake of simplicity). For simplicity, we consider the case of a single sequence, i.e., $M=1$. 
However, the approach below can be generalized to $M>1$ straightforwardly. For $M=1$, the APP of the outer code symbol $w_i$ can be computed as
 $p(w_i|\y) = \frac{p(w_i,\y)}{p(\y)}$. %
The joint probability $p(w_i,\y)$ can be computed by marginalizing the trellis states of the  inner code that correspond to symbol $w_i$. Introducing the joint state variable $\sigma_i = (s_i,d_i)$, where $s_i$ denotes the memory state variables of the convolutional code, we obtain
\begin{align*}
	p(w_i,\y) = \sum_{(\sigma,\sigma'):w_i} p(\y,\sigma,\sigma')\,,
\end{align*}
where $\sigma$ and $\sigma'$ denote realizations of the random variables $\sigma_{i-1}$ and $\sigma_i$, respectively. The summation is over all the inner code memory states that correspond to information symbol $w_i$. Introducing a drift random variable retains the Markov property of the hidden Markov model (HMM) that was lost due to the insertions and deletions \cite{davey_reliable_2001}. In this new HMM, a transition from time $i-1$ to time $i$ corresponds to a transmission of a vector of symbols $\x_{(i-1)n+1}^{in}$, where $\x_a^b = (x_a,x_{a+1},\dots, x_b)$. Further, when transitioning from state $d_{i-1}$ to $d_{i}$, the HMM emits $n+d_{i}-d_{i-1}$ output symbols depending on both the previous and the new drift. As a result, using the Markov property of the underlying trellis, we can factor the joint probability $p(\y,\sigma,\sigma')$ into three terms as
\begin{align*}
	&p(\y,\!\sigma\!,\!\sigma') \!=\!
	p\!\left(\!\y_{1}^{(i\!-\!1)n+d}, \sigma\!\right)\!p\!\left(\!\y_{(i\!-\!1)n\!+\!d+1}^{in+d'}, \sigma'\big|\sigma\!\right)\!p\!\left(\!\y_{in\!+\!d'\!+1}^{\len'}\Big| \sigma'\!\right)\!.
\end{align*}
Abbreviating the above terms by $\alpha_{i-1}(\sigma)$, $\gamma_i(\sigma,\sigma')$, and $\beta_i(\sigma')$ in order of appearance, one can deduce the  forward and backward recursions
\begin{align}
	\alpha_i(\sigma') &= \sum_{\sigma}\alpha_{i-1}(\sigma) \gamma_{i}(\sigma,\sigma')\,,  \label{eq:alpha_rec}\\
	\beta_{i-1}(\sigma) &= \sum_{\sigma'}\beta_i(\sigma') \gamma_i(\sigma,\sigma')\,, \label{eq:beta_rec}
\end{align}
where $\gamma_i(\sigma,\sigma')=p(w_i)p(\y_{(i-1)n+d+1}^{in+d'}, d'\big|d,s,s')$ can be efficiently computed using a lattice implementation \cite{bahl_decoding_1975}.

Now, $\log_2 p(\y)$ and $\log_2 p(\w,\y)$ in \eqref{eq:MIrate} can be computed based on the forward recursion in \eqref{eq:alpha_rec}. In particular, %
\begin{align*}
	p(\y) = \sum_{\sigma} p \bigl(\y_{1}^{(\leno+m)n+d}, \sigma \bigr) \overset{(a)}{=} \sum_{\sigma} \alpha_{\leno + m}(\sigma)\,,
\end{align*}
where $(a)$ follows since $\alpha_{i}(\sigma) = p\bigl(\y_{1}^{in+d}, \sigma \bigr)$. 
The quantity $\log_2 p(\w,\y)$ can be computed  in a similar manner by restricting the summation in \eqref{eq:alpha_rec} to be over all states $\sigma$  with an outgoing edge to $\sigma'$ labeled with the input sequence symbol $w_i$ at time $i$.
Since we consider an input sequence of independent and uniformly distributed symbols,  the first term $\log_2 p(\w)$ in \eqref{eq:MIrate} is  equal to $\leno\log_2 \outq$. Note that the backward recursion in \eqref{eq:beta_rec} is not required for the computation of the information density, but only for the calculation of the APP $p(w_i|\y)$ in decoding.

To obtain an estimate of the right-hand-side of \eqref{eq:BoundPf}, we randomly generate $\w$ and encode it using the inner code to obtain $\x$. Then, we pass $\x$ through the DNA storage channel to obtain $\y$. For each tuple $(\w,\y)$, we evaluate $\id(\w;\y)$ using the defined recursions and the corresponding summand in \eqref{eq:BoundPf2}. We repeat this procedure $\V$ times, each time creating a new random $\w$, and  average over the outcomes according to \eqref{eq:BoundPf2}.

\section{Concatenated Coding Scheme Design} \label{sec:Conc_Code_Scheme}
\subsection{Inner Code}
We consider four different inner codes: the watermark code  introduced in \cite{davey_reliable_2001}, a convolutional code \cite{mansour_convolutional_2010}, and  two  time-varying codes (TVCs) recently introduced in \cite{maarouf2021concatenated}. The watermark code is  an $[\leni,\dimi, 1]_q$ block code to which  a random sequence is added, which can  also be thought of as a TVC with $\tv = 1$. We will use the TVCs from \cite[Tab.~I]{maarouf2021concatenated} with $t = 4$ and a minimum  Levenshtein distance of $4$. The inner coding schemes that we  consider are summarized in Table~\ref{tab:Inner code selection}.
\begin{table}[t]
	\setlength{\tabcolsep}{1.7pt}
	\begin{center}
		\caption{Inner Synchronization Code Scheme Selection}
		\vspace{-3ex}
		{\renewcommand{\arraystretch}{1.1}
			\begin{tabular}{clccc} \specialrule{1.2pt}{0pt}{0pt}
				Scheme & Inner code & Gen. polynomial & Alt. pattern & Rate \\ \specialrule{.8pt}{0pt}{0pt}
				CC & $(1,1,2)_4$ Conv. code with RS & $[5,7]_{\text{oct}}$ & - & $0.98$ \\
				WM & $[4,4,1]_4$ Watermark code & - & - & $1.0$ \\
				TVC-$1$ & $[4,4,4]_4$ TVC & - & Random* & $1.0$ \\ 
				TVC-$2$ & $[4,4,4]_4$ TVC with RS & - & CB1 to CB4* & $1.0$ \\
				\specialrule{1.2pt}{0pt}{4pt}
			\end{tabular}\label{tab:Inner code selection}}
	\end{center}
	*The alternating pattern of the TVC-$1$ scheme is done by choosing randomly the $4$ codebooks, denoted by CB1-CB4, from \cite[Tab.~I]{maarouf2021concatenated} and avoiding consecutive codebooks. For the TVC-$2$ scheme, it is simply done by repeating CB1 to CB4 in a round Robin fashion. RS is shorthand for random sequence. 
	\vspace{-.4cm}
\end{table}
\subsection{Outer Code}

We  use protograph-based LDPC codes for the  outer code. Formally, the protograph of an LDPC code is a multi-edge-type graph with $n_{\mathsf{p}}$ variable-node (VN) types and $r_{\mathsf{p}}$ check-node (CN) types. 
A protograph can be represented by a base matrix 
\begin{align*}
    \boldsymbol{B} = \begin{pmatrix}
    b_{0,0} & b_{0,1} & \dots & b_{0,n_{\mathsf{p}}-1} \\
    b_{1,0} & b_{1,1} & \dots & b_{1,n_{\mathsf{p}}-1} \\
    \vdots & \vdots & \dots & \vdots \\
     b_{r_{\mathsf{p}}-1,0} & b_{r_{\mathsf{p}}-1,1} & \dots & b_{r_{\mathsf{p}}-1,n_{\mathsf{p}}-1}\\
    \end{pmatrix}\,,
\end{align*}
where  $b_{i,j}$ is an integer representing the number of edge connections between a type-$i$ VN and a type-$j$ CN. A parity-check matrix $\boldsymbol{H}$ of an LDPC code can be constructed from a protograph by lifting the base matrix $\boldsymbol{B}$. Lifting is the procedure of replacing each nonzero (zero) $b_{i,j}$  with a $Q_{\mathsf{p}} \times Q_{\mathsf{p}}$ permutation (zero) matrix with row and column weight equal to $b_{i,j}$.  The LDPC code resulting from the lifting procedure has length $Q_{\mathsf{p}}  n_{\mathsf{p}}$ and dimension at least $Q_{\mathsf{p}}(n_{\mathsf{p}}-r_{\mathsf{p}})$. To construct a nonbinary code from the lifted matrix, we randomly assign nonzero entries from $\field{\outq}$ to the edges of the corresponding Tanner graph.

In this work, we optimize the  protograph $\boldsymbol{B}$ using  EXIT charts, extended to the DNA storage channel. Particularly, we optimize the protograph for the case of iterations between the decoder of the LDPC code and the decoder of the combination of the inner code and the DNA storage channel.
We limit our search to protographs of dimensions $3 \times 6$ (larger protographs may lead to better performance). The choice of the protograph is done by considering both the iterative decoding threshold from the EXIT chart, for $p_\I = p_\D$ and $p_\S=0$, and the frame error rate (FER) performance of the corresponding code ensemble (i.e., by using random permutation matrices for the protograph liftings). More precisely, we sort the protographs from highest to lowest decoding threshold, and then we  pick the first protograph (starting from the top of the list) that shows no sign of an error floor above a FER of $10^{-3}$. The best protographs from this list are 
\begin{align}
\label{eq:protograph}
\bm B_1 = \left( \begin{matrix} 1 & 1 & 0 & 0 & 0 & 3 \\ 0 & 1 & 1 & 2 & 1 & 0 \\ 1 & 1 & 1 & 0 & 1 & 1 \end{matrix}\right), \bm B_2 = \left( \begin{matrix} 0 & 1 & 1 & 1 & 1 & 1 \\ 1 & 1 & 1 & 1 & 1 & 1 \\ 1 & 0 & 1 & 1 & 0 & 0 \end{matrix}\right)
\end{align}
for the CC and WM, and TVC-$1$ and TVC-$2$ inner coding schemes, respectively. We remark that the search  provided protographs with a better  threshold, but they all showed a higher error floor than $\bm B_1$ and $\bm B_2$.  All protographs were optimized for the case of $M = 1$ and over  $\field{16}$, except for the CC inner coding scheme for which $\field{2}$ was used. %

\section{Numerical Results}\label{sec:simresults}

In this section, we evaluate the DT bound \eqref{eq:BoundPf2}  %
with the inner synchronization codes listed in Table~\ref{tab:Inner code selection}. %

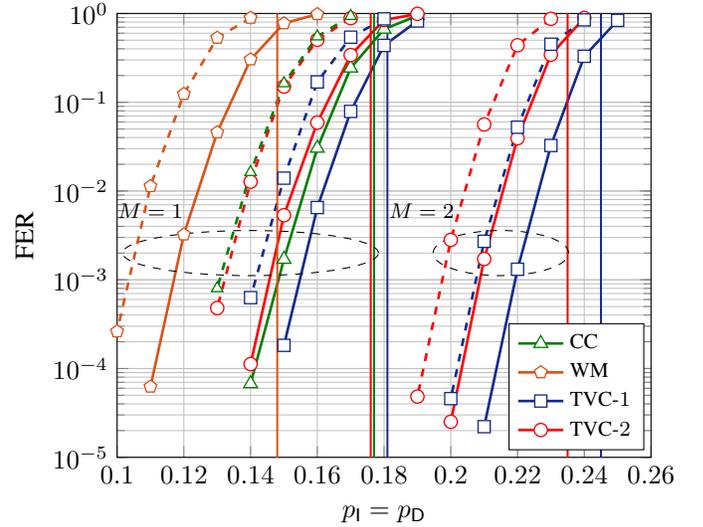
\begin{figure}[t]
    \centering
    \begin{tikzpicture}[scale=1]
\begin{semilogyaxis}[
width = 0.98\columnwidth,
xmin=0.1,   xmax=0.26,
ymin=1e-5,	ymax=1,
xticklabel style = {/pgf/number format/fixed, /pgf/number format/precision=6},
ymode=log,
grid = both,
grid style = {line width=.1pt},
legend cell align={left},
legend style={font=\footnotesize,at={(axis cs: 0.258,1.2E-5)},anchor=south east},
xlabel = {$p_\I=p_\D$},
ylabel = {FER},
cycle list name=color list
]

\addplot [mark=triangle*, line width=1.0pt, color=ColorIGMDDSR, mark options={solid, line width=0.5pt, fill=white, mark size=3pt}] table[x=pid,y=DT] {Figures/Long/DT_bound_CC_DNAchannel_N962.txt};
\addlegendentry{CC};

\addplot [mark=pentagon*, line width=1.0pt, color=ColoriBDD(ideal), mark options={solid, line width=0.5pt, fill=white, mark size=2.5pt}] table[x=pid,y=DT] {Figures/Long/DT_bound_8,4_WM_DNAchannel_N960.txt};
\addlegendentry{WM};

\addplot [mark=square*, line width=1.0pt, color=mycolor4, mark options={solid, line width = 0.5pt, fill=white, mark size=2.3pt}] table[x=pid,y=DT] {Figures/Long/DT_bound_8,4_TVC_DNAchannel_N960.txt};
\addlegendentry{TVC-$1$};

\addplot [mark=*, color=red, line width=1pt, mark options={solid, line width=0.5pt, fill=white, mark size=2.5pt}] table[x=pid,y=DT] {Figures/Long/DT_bound_8,4_TVC+RS_DNAchannel_N960.txt};
\addlegendentry{TVC-$2$};

\addplot [dashed, line width=1pt, mark=square*, color=mycolor4, mark options={solid, line width=0.5pt, fill=white, mark size=2.3pt}] table[x=pins,y=FER] {Figures/Long/Davey_MacKay_Construction_NonBinary_3,6_LDPC_FFTSPA_DNA_alphabet_BERFER_n240_k120_dv3_dc6_GF16_TVC_nw8_kw4_it1000.txt};

\addplot [dashed, mark=*, line width=1pt, color=red, mark options={solid, line width=0.5pt, fill=white,mark size=2.5pt}] table[x=pins,y=FER] {Figures/Long/Davey_MacKay_Construction_NonBinary_3,6_LDPC_FFTSPA_DNA_alphabet_BERFER_n240_k120_dv3_dc6_GF16_TVC+RS_nw8_kw4_it1000.txt};

\addplot [solid, line width= 1pt, mark=square*, color=mycolor4, mark options={solid, line width = 0.5pt, fill=white, mark size=2.3pt}] table[x=pid,y=DT] {Figures/Long/DT_bound_8,4_TVC_DNAchannel_M2_N960.txt};

\addplot [solid, line width=1pt, mark=*, color=red, mark options={solid, line width=0.5pt, fill=white, mark size=2.5pt}] table[x=pid,y=DT] {Figures/Long/DT_bound_8,4_TVC+RS_DNAchannel_M2_N960.txt};

\addplot [dashed, mark=*, line width=1pt, color=red, mark options={solid, line width=0.5pt, fill=white,mark size=2.5pt}] table[x=pins,y=FER] {Figures/Long/Davey_MacKay_Construction_NonBinary_3,6_LDPC_FFTSPA_DNA_alphabet_BERFER_n240_k120_dv3_dc6_GF16_TVC+RS_nw8_kw4_it100_M2.txt};

\addplot [dashed, line width= 1pt, mark=square*, color=mycolor4, mark options={solid, line width = 0.5pt, fill=white, mark size=2.3pt}] table[x=pins,y=FER] {Figures/Long/Davey_MacKay_Construction_NonBinary_3,6_LDPC_FFTSPA_DNA_alphabet_BERFER_n240_k120_dv3_dc6_GF16_TVC_nw8_kw4_it100_M2.txt};

\addplot [mark=pentagon*, dashed, line width=1.0pt, color=ColoriBDD(ideal), mark options={solid, line width=0.5pt, fill=white, mark size=2.5pt}] table[x=pins,y=FER] {Figures/Long/Davey_MacKay_Construction_NonBinary_3,6_LDPC_FFTSPA_DNA_alphabet_BERFER_n240_k120_dv4_dc5_GF16_WM_nw8_kw4_it100.txt};

\addplot [mark=triangle*, dashed, line width=1.0pt, color=ColorIGMDDSR, mark options={solid, line width=0.5pt, fill=white, mark size=2.5pt}] table[x=pins,y=FER] {Figures/Long/ConvolutionalCode+OptLDPC_Synchronization_DNA_alphabet_BERFER_n1924_k962_v2.txt};

\draw [color = ColorIGMDDSR, line width=0.75] (axis cs: 0.177,1) -- (axis cs: 0.177,0.00000001);
\draw [ line width=0.75, color = red] (axis cs: 0.176,1) -- (axis cs: 0.176,0.00000001);
\draw [ line width=0.75pt, color = ColoriBDD(ideal)] (axis cs: 0.148,1) -- (axis cs: 0.148,0.00000001);

\draw [ line width=0.75pt, color = mycolor4] (axis cs: 0.181,1) -- (axis cs: 0.181,0.00000001);

\draw [solid, line width=0.75pt, color = red] (axis cs: 0.235,1) -- (axis cs: 0.235,0.00000001);
\draw [solid, line width=0.75pt, color = mycolor4] (axis cs: 0.245,1) -- (axis cs: 0.245,0.00000001);

\draw[black, dashed] (axis cs: 0.14,2e-3) ellipse (1.7cm and 0.3cm) node at (axis cs: 0.11,6e-3) {\footnotesize $M = 1$};

\draw[black, dashed] (axis cs: 0.215,2e-3) ellipse (0.9cm and 0.3cm) node at (axis cs: 0.19125,6e-3) {\footnotesize $M = 2$};

\end{semilogyaxis}
\end{tikzpicture}
     \vspace{-3ex}    \caption{DT bounds (solid lines with markers)  for different inner  synchronization codes, $N=960$ DNA symbols, and $M=1$ and $M=2$. The simulated FER performance (dashed lines with markers) are for a concatenated code with an optimized outer LDPC code of rate $R = \nicefrac{1}{2}$.}
    \label{fig:FL_bounds_long}
    \vspace{-2ex}
\end{figure}

\subsection{Simulation Parameters}

We perform our simulations over the DNA alphabet $\{ \mathsf{A}, \mathsf{C}, \mathsf{G}, \mathsf{T} \}$, which corresponds to $q=4$. We consider the DNA storage channel in Figs.~\ref{fig:ids:channel} and~\ref{fig:system:model}
with $p_\S = 0$ and $p_\I=p_\D$ so that the drift random variable has zero mean (however, we remark that similar results are observed for other values and $p_\S\neq 0$). To limit the complexity of the  decoder of the combination of the inner code and the DNA storage channel, we set the maximum number of consecutive insertions considered by the decoder to $2$. Furthermore, we set the limit of the drift random variable  to five times the standard deviation of the final drift at position $\len$, i.e., to $5 \sqrt{\vphantom{A}\smash{N \frac{p_\D}{1-p_\D}}}$. Note, however, that the simulated channel may introduce more than two consecutive insertions and lead to a larger drift. 
The outer LDPC code is decoded with belief propagation with a maximum number of $100$ iterations, and the maximum number of turbo iterations is set to $100$.

We compute the DT bound for two code lengths, $N=960$ and $N=128$ DNA symbols, corresponding to a short and a medium-length sequence, respectively, and for $M = 1$ and $M = 2$ reads.
The choice of these lengths is motivated by the current  DNA sequencing technologies. All inner codes are of rate (or close to) $\ratei = 1$ (in bits per DNA symbol) and all outer codes are of rate $\rateo = \nicefrac{1}{2}$.  

\subsection{Discussion}

In Figs.~\ref{fig:FL_bounds_long} and~\ref{fig:FL_bounds_short}, we plot the DT bound (solid lines with markers) for the DNA storage channel with the inner synchronization codes in Table~\ref{tab:Inner code selection} for $N=960$ and $N=128$, respectively. The bound for $M = 2$   is obtained by considering the  \emph{joint decoding} algorithm proposed  in~\cite{maarouf2021concatenated}. Furthermore, in the figures we plot the asymptotic achievable information rates (vertical lines) computed in \cite{maarouf2021concatenated} for each inner coding scheme. 

The TVC-$1$ scheme yields the best bound for both code lengths and values of $M$, and the watermark code gives the worst bound. Interestingly, the hierarchy of the bounds coincides with the hierarchy of the asymptotic achievable information rates.
\begin{figure}[t]
    \centering
    \begin{tikzpicture}[scale=1]
\begin{semilogyaxis}[
width = 0.98\columnwidth,
xmin=0.05,   xmax=0.25,
ymin=1e-5,	ymax=1,
xticklabel style = {/pgf/number format/fixed, /pgf/number format/precision=6},
ymode=log,
grid = both,
grid style = {line width=.1pt},
legend cell align={left},
legend style={font=\footnotesize,at={(axis cs: 0.245,1.5E-5)},anchor=south east},
xlabel = {$p_\I=p_\D$},
ylabel = {FER},
cycle list name=color list
]

\addplot [mark=square*, line width=1.0pt, color=mycolor4, mark options={solid, line width = 0.5pt, fill=white, mark size=2.3pt}] table[x=pid,y=DT] {Figures/Short/DT_bound_8,4_TVC_DNAchannel_N128.txt};
\addlegendentry{TVC-$1$};

\addplot [mark=*, color=red, line width=1pt, mark options={solid, line width=0.5pt, fill=white, mark size=2.5pt}] table[x=pid,y=DT] {Figures/Short/DT_bound_8,4_TVC+RS_DNAchannel_N128.txt};
\addlegendentry{TVC-$2$};

\addplot [dashdotted, mark=square*, line width=1.0pt, color=mycolor4, mark options={solid, line width = 0.5pt, fill=white, mark size=2.3pt}] table[x=pins,y=FER] {Figures/Short/Davey_MacKay_Construction_NonBinary_3,6_LDPC_FFTSPA_DNA_alphabet_BERFER_n30_k15_dv3_dc6_GF16_TVC_nw8_kw4_it1000.txt};

\addplot [dashdotted, mark=*, color=red, line width=1pt, mark options={solid, line width=0.5pt, fill=white, mark size=2.5pt}] table[x=pins,y=FER] {Figures/Short/Davey_MacKay_Construction_NonBinary_3,6_LDPC_FFTSPA_DNA_alphabet_BERFER_n30_k15_dv3_dc6_GF16_TVC+RS_nw8_kw4_it1000.txt};

\addplot [mark=square*, line width=1.0pt, color=mycolor4, mark options={solid, line width = 0.5pt, fill=white, mark size=2.3pt}] table[x=pid,y=DT] {Figures/Short/DT_bound_8,4_TVC_DNAchannel_M2_N128.txt};

\addplot [mark=*, color=red, line width=1pt, mark options={solid, line width=0.5pt, fill=white, mark size=2.5pt}] table[x=pid,y=DT] {Figures/Short/DT_bound_8,4_TVC+RS_DNAchannel_M2_N128.txt};

\addplot [dashdotted, mark=*, color=red, line width=1pt, mark options={solid, line width=0.5pt, fill=white, mark size=2.5pt}] table[x=pins,y=FER] {Figures/Short/Davey_MacKay_Construction_NonBinary_3,6_LDPC_FFTSPA_DNA_alphabet_BERFER_n30_k15_dv3_dc6_GF16_TVC+RS_nw8_kw4_it100_M2.txt};

\addplot [dashdotted, mark=square*, line width=1.0pt, color=mycolor4, mark options={solid, line width = 0.5pt, fill=white, mark size=2.3pt}] table[x=pins,y=FER] {Figures/Short/Davey_MacKay_Construction_NonBinary_3,6_LDPC_FFTSPA_DNA_alphabet_BERFER_n30_k15_dv3_dc6_GF16_TVC_nw8_kw4_it100_M2.txt};

\draw [solid, line width=0.75pt, color = red] (axis cs: 0.176,1) -- (axis cs: 0.176,0.00000001);

\draw [ solid, line width=0.75pt, color = mycolor4] (axis cs: 0.181,1) -- (axis cs: 0.181,0.00000001);

\draw [solid, line width=0.75pt, color = red] (axis cs: 0.235,1) -- (axis cs: 0.235,0.00000001);
\draw [solid, line width=0.75pt, color = mycolor4] (axis cs: 0.245,1) -- (axis cs: 0.245,0.00000001);

\draw[black, dashed] (axis cs: 0.105,2e-3) ellipse (0.7cm and 0.3cm) node at (axis cs: 0.124,7e-4) {\footnotesize $M = 1$};

\draw[black, dashed] (axis cs: 0.18,2e-3) ellipse (0.7cm and 0.3cm) node at (axis cs: 0.198,7e-4) {\footnotesize $M = 2$};

\end{semilogyaxis}%
\end{tikzpicture}%
    \vspace{-4ex}
    \caption{DT bounds (solid lines with markers)  for the TVC-$1$ and TVC-$2$ inner  coding schemes, $N=128$ DNA symbols, and $M=1$ and $M=2$. The simulated FER performance (dashed lines with markers) are for a concatenated code with an optimized outer LDPC code of rate $R = \nicefrac{1}{2}$.}
    \label{fig:FL_bounds_short}
    \vspace{-2ex}
\end{figure}
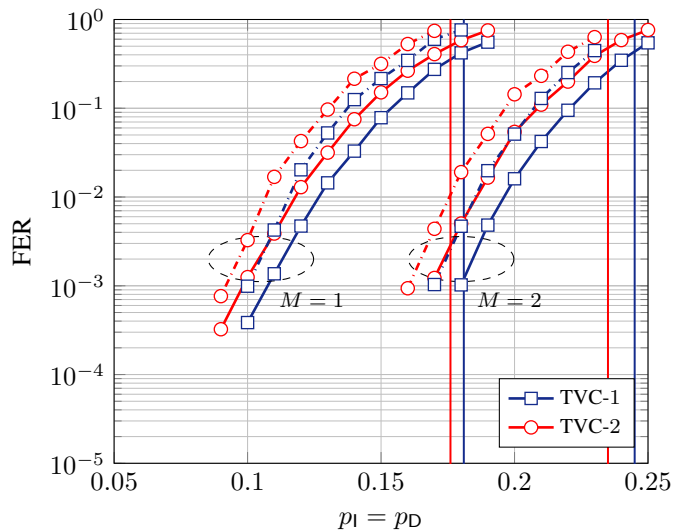
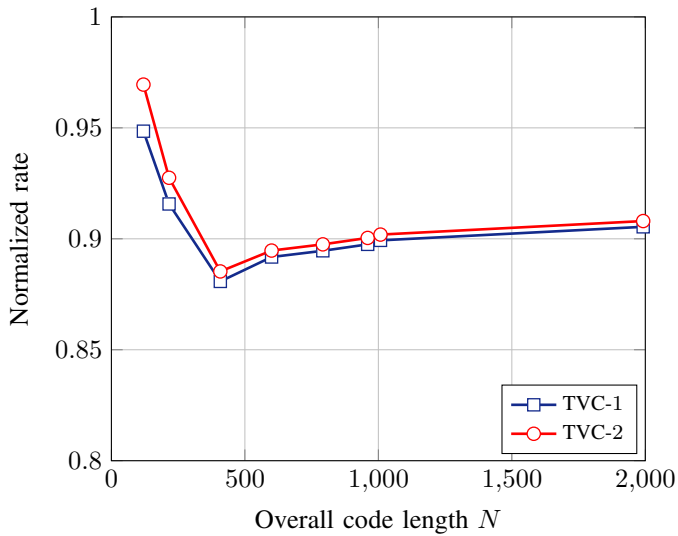
\begin{figure}[t]
    \centering
    \begin{tikzpicture}[scale=1]
\begin{axis}[
width = 0.98\columnwidth,
xmin=0,   xmax=2000,
ymin=0.8,	ymax=1,
xticklabel style = {/pgf/number format/fixed, /pgf/number format/precision=20},
grid = both,
grid style = {line width=.2pt},
legend cell align={left},
legend style={font=\footnotesize,at={(axis cs: 1970,0.803)},anchor=south east},
xlabel = {Overall code length $\len$},
ylabel = {Normalized rate},
cycle list name=color list
]

\addplot [mark=square*, line width=1.0pt, color=mycolor4, mark options={solid, line width = 0.5pt, fill=white, mark size=2.3pt}] table[x=l,y=r] {Figures/Relative_to_rate_performance/TVC_DTbound2.txt};
\addlegendentry{TVC-$1$}

\addplot [mark=*, color=red, line width=1pt, mark options={solid, line width=0.5pt, fill=white, mark size=2.5pt}] table[x=l,y=r] {Figures/Relative_to_rate_performance/TVC+RS_DTbound2.txt};
\addlegendentry{TVC-$2$}

\end{axis}
\end{tikzpicture}%
     \vspace{-4ex}
    \caption{Normalized rate for a concatenated coding scheme with an optimized outer LDPC code    constructed from the protograph $\bm B_2$ in \eqref{eq:protograph} and the TVC-$1$ and TVC-$2$ inner  coding schemes as a function of the  code length $\len$. The overall code rate is  $R = \nicefrac{1}{2}$ and the target FER is $10^{-3}$.}
    \label{fig:Relative_to_rate}
    \vspace{-2ex}
\end{figure}

In the figures, we also plot the FER performance (dashed lines with markers) for a concatenated code with  an outer LDPC code built from the optimized protographs in \eqref{eq:protograph} and  the inner  coding schemes from Table~\ref{tab:Inner code selection}. In contrast to the optimization, circulant matrices for the protograph liftings,  built using the progressive edge-growth algorithm~\cite{PEG}, are used. The slope of the FER curves is similar to the slope of the corresponding DT bounds and a similar gap to the bounds is observed for the simulated FER curves. Notably, the proposed concatenated schemes perform close to the DT bounds. 

To gain more insight on  the  performance of the proposed concatenated schemes to the DT bound, in  Fig.~\ref{fig:Relative_to_rate} we plot the normalized rate  \cite{DT+RCUsbounds} as a function of the code length $N$ for the concatenated code with the TVC-$1$ and TVC-$2$ inner  coding schemes. The normalized rate is computed as the fraction between the rate of the concatenated code and the maximum rate provided by the DT bound so that decoding with a probability of error below a given value is possible.  %
In other words, we want a normalized rate close to one and a normalized rate of one means that the code achieves the DT bound. In the plot, we consider a FER of $10^{-3}$.

For both TVC-$1$ and TVC-$2$, the normalized rate is within $87\%$ to $97\%$ for a code length up to $\len=2000$ DNA symbols. These values are similar to those for state-of-the-art codes over memoryless channels \cite[Fig.~15]{DT+RCUsbounds}, indicating that the proposed concatenated codes yield excellent performance on the DNA storage channel.

\section{Conclusion}
\label{sec:future_work} 
We provided an upper bound to the performance of random coding schemes on a DNA storage channel with insertions, deletions, and substitutions in the practical short-to-medium blocklength regime. The bound, which is based on the dependency testing bound yields an achievability result and is particularly useful to capture the performance of  concatenated coding schemes with an inner synchronization code as it depends on the inner code. Hence, it is  a handy tool to guide the choice of the inner synchronization code and provides a reference to benchmark the performance of coding schemes. 

\ifCLASSOPTIONcaptionsoff
  \newpage
\fi
%


\end{document}